\documentclass[12pt]{article}
\usepackage{a4}

\usepackage{amsmath}
\usepackage{amsfonts}
\usepackage{amsmath}
\usepackage{latexsym}
\usepackage{amsfonts}
\begin{document}
\oddsidemargin 6pt\evensidemargin 6pt\marginparwidth
48pt\marginparsep 10pt

\renewcommand{\thefootnote}{\fnsymbol{footnote}}
\thispagestyle{empty}

%\noindent    \hfill  June 2007 \\

\noindent \vskip3.3cm
\begin{center}

{\Large\bf Conformal invariant powers of the Laplacian, Fefferman-Graham ambient metric and Ricci gauging}
\bigskip\bigskip\bigskip

{\large Ruben Manvelyan, Karapet Mkrtchyan and Ruben Mkrtchyan}\\
\medskip

{\small\it Department of Theoretical Physics\\ Yerevan Physics Institute\\ Alikhanian Br.
Str.
2, 0036 Yerevan, Armenia}\\
\medskip
{\small\tt e-mails manvel@physik.uni-kl.de, karapet@yerphi.am, mrl@web.am}
\end{center}

\bigskip %\bigskip
\begin{center}
{\sc Abstract}
\end{center}

The hierarchy of conformally invariant $k$-th powers of the Laplacian acting
on a scalar field with scaling dimensions $\Delta_{(k)}=k-d/2$, $k=1,2,3$
as obtained in the recent work \cite{MT}  is rederived using the Fefferman-Graham $d+2$ dimensional ambient space approach. The corresponding mysterious "holographic" structure of these operators is clarified. We explore also the $d+2$ dimensional ambient space origin of the Ricci gauging procedure proposed by A.~Iorio, L.~O'Raifeartaigh, I.~Sachs and C.~Wiesendanger as another method of constructing the Weyl invariant Lagrangians.
The corresponding \emph{gauged} ambient metric, Fefferman-Graham expansion  and extended
Penrose-Brown-Henneaux transformations are proposed and analyzed.

\newpage

\section{Introduction}

\quad
The problem of constructing conformally invariant Lagrangians or differential operators in various dimensions and for various fields has quite a long history. This problem attracts attention primarily because it is always a nontrivial task to construct local conformal or Weyl invariants in higher dimensions \cite{Bon, Erd, Boul}. The $AdS/CFT$ correspondence \cite{mald} increased interest in this old problem as well as returned the attention to the seminal mathematical paper by Fefferman and Graham (FG) on conformal invariants  \cite{FG}.
Our goal in this article is to establish the connections between different ways of construction of the local conformal invariant Lagrangians or differential operators in $d$ dimensions \cite{MT},\cite{oraf} and the FG $d+2$ dimensional ambient  Ricci flat space method \cite{FG}.
In other words we extend the consideration of the article \cite{FG} to reproduce and explain the results obtained in \cite{MT} and \cite{oraf}. The main FG idea consist in the confidence that the lower dimensional diffeomorphisms and local conformal invariants can be obtained from corresponding reparametrization invariant counterparts in the higher dimensional space where $d$ dimensional conformal invariance is realized as a part of $d+2$ dimensional diffeomorphisms (we review the FG method in Section 2). On the other hand the FG expansion is connected with $AdS_{d+1}/CFT_{d}$ correspondence and plays a crucial role in derivation of the holographic anomalies in different dimensions \cite{HS}. This point forced us in Section 3 to derive again, using the FG ambient space method, the
hierarchy of conformally invariant powers of the Laplacian (or invariant Lagrangian) in spacetime dimensions $d\geq 2k$ acting on a scalar field obtained in \cite{MT} by the direct Noether procedure, whose conformal dimension is $\Delta_{(k)}=k-d/2$. This ambient space derivation unveiled the remarkable and mysterious feature of these differential invariants namely the appearance of the $2k$ dimensional holographic anomaly in the $k$-th member of this hierarchy \cite{MT} (the most recent mathematical development in the holographic  formalism for conformally invariant operators is considered in \cite{Juhl}).

In this article we propose also (Section 4) an extended or gauged FG $d+2$ dimensional space to establish a connection  between  the FG expansion and another interesting method of constructing  the Weyl invariant Lagrangians obtained in \cite{oraf} by A.~Iorio, L.~O'Raifeartaigh, I.~Sachs and C.~Wiesendanger and  named ``Ricci gauging''. The magic and universality of the $d+2$ dimensional FG method is defined by the existence of so-called Penrose-Brown-Henneaux (PBH) diffeomorphisms \cite{PBH} considered in details for usual FG metric in \cite{Schwim} and \cite{OA}. In section 3 we consider the new PBH transformation for gauged ambient spaces to explore some properties of the FG expansion in the presence of the Weyl gauge field and the holographic origin of the Ricci gauging.

\section{Ambient metric and Fefferman-Graham\\ expansion}
In this section we review the FG ambient space method for constructing local conformal invariants \cite{FG}.
We define the $d+2$ dimensional ambient space with the set of coordinates
$\{x^{\mu}\}= \{t,\rho,x^{i}; i=1,2,\dots d \}$ and the following Ricci flat
\footnote{We use the same conventions for covariant derivatives and curvatures as in \cite{MT}:
\begin{eqnarray}
  &&\nabla_{\mu}V^{\rho}_{\lambda}= \partial_{\mu}V^{\rho}_{\lambda}+
  \Gamma^{\rho}_{\mu\sigma}V^{\sigma}_{\lambda}-\Gamma^{\sigma}_{\mu\lambda}V^{\rho}_{\sigma} , \quad\Gamma^{\rho}_{\mu\nu} = \frac{1}{2} g^{\rho\lambda}\left(\partial_{\mu}g_{\nu\lambda}+
  \partial_{\nu}g_{\mu\lambda} - \partial_{\lambda}g_{\mu\nu}\right) ,\nonumber\\
  &&\left[\nabla_{\mu} , \nabla_{\nu}\right]V^{\rho}_{\lambda} =
  R^{\quad\,\,\rho}_{\mu\nu \sigma}V^{\sigma}_{\lambda}
  -R^{\quad\,\,\sigma}_{\mu\nu \lambda}V^{\rho}_{\sigma} ,\quad
  R^{\quad\,\,\rho}_{\mu\nu \lambda}= \partial_{\mu}\Gamma^{\rho}_{\nu\lambda}
  -\partial_{\nu}\Gamma^{\rho}_{\mu\lambda}+\Gamma^{\rho}_{\mu\sigma}\Gamma^{\sigma}_{\nu\lambda}
  -\Gamma^{\rho}_{\nu\sigma}\Gamma^{\sigma}_{\mu\lambda} ,\nonumber\\
  &&R_{\mu\lambda}= R^{\quad\,\,\rho}_{\mu\rho\lambda}\quad , \quad
  R=R^{\,\,\mu}_{\mu} .\nonumber
\end{eqnarray}} metric
\begin{equation}\label{1}
ds^{2}_{A}=g^{A}_{\mu\nu}(t,\rho,x^{i})dx^{\mu}dx^{\nu}=
\frac{t^{2}}{\ell^{2}}h_{ij}(x,\rho)dx^{i}dx^{j}-\rho dt^{2}-tdtd\rho\,,
\end{equation}
where
\begin{equation}\label{2}
    h_{ij}(x,\rho)=g_{ij}(x)+\rho h^{(1)}_{ij}(x)+\rho^{2}h^{(2)}_{ij}(x)+\dots\dots
\end{equation}
is the well known FG expansion with an arbitrary boundary value of the metric $g_{ij}(x)=h_{ij}(x,\rho)|_{\rho=0}$ and  a set of  the higher $\rho$ derivatives $n!h^{(n)}_{ij}(x)=\frac{\partial^{n}}{\partial\rho^{n}}h_{ij}(x,\rho)|_{\rho=0}$ fixed by the Ricci flatness condition in ambient space
\begin{equation}\label{3}
    R^{A}_{\mu\nu}=0\,.
\end{equation}
This condition produces the following set of equations
\begin{eqnarray}
  && R^{A}_{it} = R^{A}_{\rho t}= R^{A}_{tt}\equiv 0 ,\label{4}\\
  &&R^{A}_{\rho\rho} =\frac{1}{2}\left[h^{kl}h''_{kl}
  -\frac{1}{2}h^{ij}h'_{jk}h^{kl}h'_{li}\right]=0 \,,\label{5}\\
  && R^{A}_{i\rho}=\frac{1}{2}h^{kl}\left[\nabla^{(h)}_{i}
  h'_{kl}-\nabla^{(h)}_{k}h'_{il}\right]=0\,,\label{6}\\
  &&\ell^{2}R^{A}_{ij}=\ell^{2}R_{ij}[h]-(d-2)h'_{ij} - h^{kl}h'_{kl}h_{ij}\nonumber\\&&+
  \rho\left[2h''_{ij}-2h'_{il}h^{lm}h'_{mj}
  +h^{kl}h'_{kl}h'_{ij}\right]=0\,,\label{7}
\end{eqnarray}
where $\dots'=\partial_{\rho}\dots$ and $\nabla^{(h)}_{i}, R_{ij}[h]$ are covariant derivative and Ricci tensor of the metric
$h_{ij}(x,\rho)$, respectively. It was shown in \cite{FG} that this system of equations is equivalent to the
$d+1$ dimensional Einstein's equations with negative cosmological constant (see \cite{OA} for details). This can be easily seen from the following consideration
\begin{itemize}
  \item The $AdS_{d+1}$ bulk can be found in $d+2$ dimensional ambient space as a $d+1$ dimensional surface defined as
      \begin{equation}\label{8}
        t^{2}\rho=\ell^{2} ,\quad \rho >0.
      \end{equation}
      on which the metric (\ref{1}) induces the standard Poincar\'{e} metric for coordinates $\{x^{a}\}=\{\rho, x^{i}\}$
      \begin{equation}\label{9}
      ds^{2}_{Bulk}=g^{Bulk}_{ab}(x,\rho)dx^{a}dx^{b}
      =\frac{\ell^{2}}{4\rho^{2}}d\rho^{2}+\frac{1}{\rho}h_{ij}(x,\rho)dx^{i}dx^{j}\,.
\end{equation}
    \item The corresponding bulk Ricci tensor is related to the nonzero components of the ambient Ricci tensor in the way
        \begin{equation}\label{10}
            R^{A}_{ab}=R^{Bulk}_{ab}+\frac{d}{\ell^{2}}g^{Bulk}_{ab}\,,
        \end{equation}
       and condition (\ref{3}) leads to the negative constant curvature
       \begin{equation}\label{11}
         R^{A}_{ab}=0 \Rightarrow R^{Bulk}=R^{Bulk}_{ab}g^{ab}_{Bulk}=-\frac{d(d+1)}{\ell^{2}}\,.
       \end{equation}
       \end{itemize}
 Therefore  (\ref{3}) leads, as in the case of $AdS_{d+1}/CFT_{d}$ correspondence \cite{HS}, to the same solutions for $h^{(n)}_{ij}(x)$ in the FG expansion (\ref{2}) in terms of covariant objects constructed from the boundary value $g_{ij}(x)$
 \begin{eqnarray}
   && h^{(1)}_{ij}(x)=\ell^{2}K_{ij} ,\quad
   h^{(1)}=g^{ij}(x)h^{(1)}_{ij}(x)=\ell^{2}J \,,\label{12}\\
   &&h^{(2)}_{ij}(x)=\frac{\ell^{4}}{4}\left\{\frac{B_{ij}}{d-4}+K^{m}_{i}K_{mj}\right\} , \quad  h^{(2)}=g^{ij}(x)h^{(2)}_{ij}(x)=\frac{\ell^{4}}{4}K^{ij}K_{ij},\label{13}\\
   &&  h^{(3)}=g^{ij}(x)h^{(3)}_{ij}(x)=\frac{\ell^{6}}{6(d-4)}K^{ij}B_{ij} \,,\label{14}
 \end{eqnarray}
where we introduced the Weyl ($W$) and Schouten ($K$) tensors, as well as
the scalar $J$  and so called Cotton ($C$) and Bach ($B$) tensors\footnote{All important properties of these tensors following from the Bianchi
identity can then be listed as (from now on  $\nabla_{i}$ is the covariant derivative of the metric $g_{ij}(x)$)
\begin{eqnarray}
  \nabla _{[m} W_{ij ]k }^{\quad\,\,\,n} &=& g_{k [m }
  C_{ij ]}^{\quad\,\,n}
   - \delta _{[m }^n  C_{ij ]k } \,\,,\,
 \nabla _m  W_{ijk}^{\quad\,\,m}= \left( {3
- d}\right)C_{ijk } \,\,\,,\nonumber\\
 \nabla^{i}K_{ij}&=&\partial_{j}J\,,\quad
\nabla^{k}C_{ijk}=0\,\,\,,\quad  C_{ij }^{\quad\,j} = 0 \,,\quad
 \nabla^{i}B_{ij}=(d-4)C_{ijk}K^{ik} .\nonumber
\end{eqnarray}}
\begin{eqnarray}
  K_{ij} &=& \frac{1}{(d - 2)}\left\{R_{ij} - g_{ij}J\right\} , \quad J=\frac{1}{2(d - 1)}R\,\,, \label{15}\\
  W_{ijk}^{\quad\,\,m} &=& R_{ijk }^{\quad\,\,m} -K_{ik}\delta^{m}_{j}  +
 K_{jk} \delta^{m}_{i} - K^{m}_{j} g_{ik}  +
 K^{m}_{i}g_{jk} \,\,, \label{16}\\
B_{ij}&=&\nabla^{k}C_{kij}
  +K^{k}_{m}W_{kij }^{\quad\,\,m}, \quad C_{ijk} = \nabla _i  K_{jk}  - \nabla _j
 K_{ik } \,.\label{17}
 \end{eqnarray}
This basis of $B,C,K,J,W$ tensors was used in \cite{MT} to construct directly\footnote{This basis of $B,C,K,J,W$ tensors forms  a closed system with respect to local conformal (or Weyl) transformations of the boundary metric $\delta g_{ij}(x)=2\sigma(x)g_{ij}(x)$
\begin{eqnarray}
 && \delta W_{ijk}^{\quad\,\,m}=0 ,\quad  \delta K_{ij}=\nabla_{i}\partial_{j}\sigma ,\quad \delta J=-2\sigma J +\Box \sigma\nonumber\\
  && \delta C_{ijk}= -\partial_{m}\sigma W_{ijk}^{\quad\,\,m} ,\quad \delta B_{ij}= -2\sigma B_{ij} +(d-4)\nabla^{k}\sigma \left(C_{kij}+C_{kji}\right), \nonumber
\end{eqnarray} and it is all one needs to construct any conformally invariant object in arbitrary dimensions \cite{Erd}, \cite{MT}.}
a hierarchy of
conformally invariant Lagrangians or differential operators originating from  powers of the Laplacian in spacetime dimensions $d\geq
2k$, describing the nonminimal coupling of gravity with a scalar
field whose conformal dimension is $\Delta_{(k)}=k-d/2$.

The connection of the Fefferman-Graham ambient metric expansion and\\ $AdS/CFT$ correspondence was investigated and developed by many authors. We do not pretend here to present an exhaustive list of citations in this field and just quote a number of  articles important for us in this area \cite{HS}, \cite{Schwim}, \cite{OA}. For us the most important result of \cite{FG} is the elegant method of constructing conformal invariants (covariants) in $d$ dimensions from reparametrization invariant (covariant) combinations of the curvature and it's covariant derivatives in $d+2$ dimensional ambient space equipped with a Ricci flat metric (\ref{1}) by truncation to the $d$ dimensional boundary  at $\rho=0$ and $t=const$. In the simplest case of a Riemannian  curvature tensor this prescription gives for nonvanishing  components  (see \cite{OA} for detailed derivation)
\begin{eqnarray}
  && R^{A\quad l}_{ijk}|_{\rho=0}=W_{ijk}^{\quad l} \,,\label{18}\\
  && R^{A\quad t}_{ijk}|_{\rho=0}=t C_{ijk} \label{19}\,,\\
  && R^{A\quad t}_{\rho ij}|_{\rho=0}=\frac{t \ell^{2}}{2}\frac{B_{ij}}{d-4}\,. \label{20}
\end{eqnarray}
Using this the authors derived in \cite{FG} the first nontrivial invariant obtained from $(\nabla^{A}_{m}R^{A}_{ijkl})^{2}$ and discussed in details in \cite{Erd}. In the same article Fefferman and Graham predicted that usual Laplacian in ambient $d+2$ dimensional space should produce conformal invariant second order differential operator in dimension $d$, which is the first representative in the hierarchy of conformal operators for scalar fields constructed in \cite{MT}.

\section{Hierarchies of conformal invariant powers of Laplacian from ambient space }

\quad In \cite{MT} the authors introduced the hierarchy of scalar fields
$\varphi_{(k)}$, where $k=1,2,3,\dots$ with the corresponding
scaling dimensions and infinitesimal conformal transformations
\begin{eqnarray}\label{trans}
   \Delta_{(k)}&=&k-d/2\,,\label{21}\\
    \delta \varphi_{(k)}(x):&=&\Delta_{(k)}\sigma(x)\varphi_{(k)}(x) .\label{22}
\end{eqnarray}
Each of these exists in the spacetime dimensions $d\geq 2k$, and with
the minimal vanishing dimension, $\Delta_{(k)}=0$ when $d=2k$ and couples with gravity in the conformally invariant way through the hierarchy of the conformally invariant $k$-th power of the Laplacian
\begin{equation}\label{23}
    \hat{\mathcal{L}}_{(k)}=\Box^{k}+\dots\dots+\Delta_{(k)}\mathbf{a}_{{(k)}} .
\end{equation}
The interesting point of the consideration in \cite{MT} was the appearance of the so-called holographic anomaly $\mathbf{a}_{{(k)}}$ \cite{HS} namely
\emph{the derivative independent part of the conformally
invariant $k$-th power of Laplacian is the scaling dimension times the holographic anomaly in dimension $d=2k$ written in general spacetime dimension $d$}.

In this section we will explain this remarkable property of the above hierarchy, namely that one obtains
conformal invariant operators from the $k$-th power of the Laplace-Beltrami operator constructed from the ambient metric which acts on the $d+2$ dimensional scalar field and from using the FG holographic expansion (\ref{2}). So we concentrate on \footnote{We use the notation  $\Box_{A}$ for the Laplacian in ambient space. The $\Box_{h}$ is the Laplacian constructed from $h_{ij}(x,\rho)$ and  a simple $\Box$ corresponds to the boundary metric $g_{ij}(x)$ .}
\begin{equation}\label{24}
    \left(\Box_{A}\right)^{k}f(x,t,\rho)\,,
\end{equation}
where
\begin{equation}\label{25}
    \Box_{A}=\frac{\ell^{2}}{t^{2}}\Box_{h}
    +\frac{4\rho}{t^{2}}\partial^{2}_{\rho}
    -\frac{4}{t}\partial_{t}\partial_{\rho}
    +h^{ij}h'_{ij}\left(\frac{2\rho}{t^{2}}\partial_{\rho}
    -\frac{1}{t}\partial_{t}\right)-\frac{2(d-2)}{t^{2}}\partial_{\rho}\,.
    \end{equation}
For doing that first of all we have to understand the right truncation for the $d+2$ dimensional scalar $f(x,t,\rho)$ to the $d$ dimensional scalar $\varphi_{k}(x)$. Taking into account that we do not want to consider $AdS/CFT$ behaviour for the scalar field we can  take it $\rho$ independent. Then from simple scaling arguments we arrive at the following ansatz
\begin{equation}\label{26}
    f(x,t,\rho) = t^{\Delta_{(k)}}\varphi_{(k)}(x)\,.
\end{equation}
Then we see that (\ref{25}) reduces to
\begin{eqnarray}
  \Box_{A}\left[t^{\Delta_{(k)}}\varphi_{(k)}(x)\right]
  &=&\ell^{2}t^{\Delta_{(k)}-2}\left[\Box_{h} \varphi_{(k)}(x)-\frac{\Delta_{(k)}}{\ell^{2}}h^{ij}h'_{ij}\varphi_{(k)}(x)\right] \,, \label{27}
\end{eqnarray}
so that inserting $k=1$ and using (\ref{12}) we obtain
\begin{equation}\label{28}
     \Box_{A}\left[t^{\Delta_{(1)}}\varphi_{(1)}(x)\right]|_{\rho=0}=\ell^{2}t^{-d/2}\left(\Box -\Delta_{(1)}J\right)\varphi_{(1)}(x) ,
\end{equation}
where we recognize in the brackets the well known conformal Laplacian
\begin{equation}\label{29}
   \hat{\mathcal{L}}_{(1)}= \Box -\Delta_{(1)}J = \Box+\frac{(d-2)}{4(d-1)}R\,.
\end{equation}

The next step in our ambient space considerations is the $k=2$ case.
First we rewrite the last term in (\ref{25}) in the $\Delta_{(k)}$ dependent form
\begin{equation}\label{30}
    -2\frac{d-2}{t^{2}}\partial_{\rho}=\frac{4\Delta_{(k)}-4(k-1)}{t^{2}}\partial_{\rho}\,.
\end{equation}
Inserting (\ref{27}) in (\ref{25}) and expanding in $\rho$ we obtain
\begin{eqnarray}
   &&\Box^{2}_{A}\left[t^{\Delta_{(k)}}\varphi_{(k)}(x)\right]
  =\ell^{4}t^{\Delta_{{(k)}}-4}f_{(k)}(\rho,x)=\ell^{4}t^{\Delta_{{(k)}}-4}\Big\{\left(\Box- \frac{\Delta_{(k)}}{\ell^{2}}h^{(1)}\right)^{2}+\frac{2}{\ell^2}h^{(1)}\Box \nonumber\\
  && -\frac{4(3-k)}{\ell^{2}}\left[h^{(1)ij}\nabla_{i}\partial_{j}+\frac{1}{2}(\nabla^{n}h^{(1)})\partial_{n}\right]
  +\frac{2}{\ell^{4}}\Delta_{(k)}\left[(3-k)h^{(1)ij}h^{(1)}_{ij}-h^{(1)^{2}}\right] \nonumber\\
  &&+\frac{\rho\Delta_{(k)}}{\ell^{4}}\left( 8h^{(1)ij}h^{(2)}_{ij}
  -4h^{(1)ij}h^{(1)}_{jn}h^{(1)n}_{i}+ 3h^{(1)}
  h^{(1)ij}h^{(1)}_{ij}\right)\nonumber\\
   && + \rho O(\nabla)+\rho O(3-k)+ \rho O(\Delta_{(k)}^{2})+O(\rho^{2})\Big\}\varphi_{(k)}(x) ,\,\,\,\,\label{31}
\end{eqnarray}
where we use the  following relations
\begin{eqnarray}
  && \nabla_{j}h^{(1)j}_{i}=\nabla_{i}h^{(1)} ,\quad h^{(2)}=\frac{1}{4}h^{(1)ij}h^{(1)}_{ij}\,, \label{32}\\
  && \nabla_{j}h^{(2)j}_{i}+\frac{1}{2}\nabla_{i}h^{(2)}
  =\frac{1}{2}h^{(1)jn}\nabla_{j}h^{(1)}_{ni}+\frac{1}{4}h^{(1)}_{ij}\nabla^{j}h^{(1)} \,, \label{33}\\
  &&h^{(3)}=\frac{2}{3}h^{(1)ij}h^{(2)}_{ij}-\frac{1}{6}h^{(1)ij}h^{(1)}_{jn}h^{(1)n}_{i} \,, \label{34}
\end{eqnarray}
obtained from $\rho$ expansion of  (\ref{5}) and (\ref{6}).
Now inserting in (\ref{31}) $k=2$ and $\rho=0$  and using (\ref{12}) we obtain
\begin{eqnarray}
  &&\Box^{2}_{A}\left[t^{\Delta_{(2)}}\varphi_{(k)}(x)\right]|_{\rho=0}
  =\ell^{4}t^{\Delta_{{(2)}}-4}\hat{\mathcal{L}}_{(2)}\varphi_{(k)}(x) \,,\label{35}\\
  && \hat{\mathcal{L}}_{(2)}=\left(\Box-\Delta_{(2)}J\right)^{2}
  -4\nabla_{i}K^{ij}\partial_{j}+2\nabla^{i}J\partial_{i}
  +2\Delta_{(2)}\left(K^{2}-J^{2}\right)\,. \label{36}
\end{eqnarray}
Again this fourth order
higher derivative  conformal invariant
operator is known since many years~\cite{Riegert,FT} for dimension
$4$ as well as for general $d$~\cite{pan,es}. This operator was rederived in \cite{MT} as a kinetic operator for the second  Lagrangian of the hierarchy of conformally coupled scalars by simply applying the Noether procedure.

Now we can evaluate the general expression for Euler densities
\begin{eqnarray}
  && E_{(k)}:=\frac{1}{2k (d-2k)!}\delta^{i_{1}
  \dots i_{d-2k}j_{1}j_{2}\dots j_{2k-1}j_{2k}}_{i_{1}
  \dots i_{d-2k}k_{1}k_{2}\dots k_{2k-1}k_{2k}}
  R^{k_{1}k_{2}}_{j_{1}j_{2}}\dots R^{k_{2k-1}k_{2k}}_{j_{2k-1}
  j_{2k}} \,.\label{37}
\end{eqnarray}
for $k=2$ and obtain
\begin{eqnarray}
  2\Delta_{(2)}\left(K^{2}-J^{2}\right)=-\frac{\Delta_{(2)}}{2(d-3)(d-2)}\left(E_{(2)}- W^{2}\right)\,. \label{38}
\end{eqnarray}
So we see that the last term in (\ref{36}), which is linear in $\Delta_{(2)}$, is proportional to the
Weyl tensor independent part of the Euler
density.  Thus we recognize as $\mathbf{a}_{(k)}$ of (\ref{23}) for both the $k=1,2$ cases (\ref{28}), (\ref{35})
\begin{eqnarray}
  && \mathbf{a}_{(1)}=-\frac{1}{\ell^{2}}h^{(1)}=-\frac{1}{2(d-1)}E_{(1)} \,,\label{39}\\
  && \mathbf{a}_{(2)}=2(h^{(1)ij}h^{(1)}_{ij}-h^{(1)2})
  =-\frac{1}{2(d-3)(d-2)}\left(E_{(2)}- W^{2}\right)\,.\label{40}
\end{eqnarray}

The "holographic" trace anomaly arises in $AdS/CFT$ \cite{HS}
and corresponds to the maximally supersymmetric gauge
theories on the boundary of $AdS_{3}$ and $AdS_{5}$.
To check our statement as an assertion for general $k$, we need to carry out this verification
in the next nontrivial case of $k=3$ obtained in \cite{MT} by the Noether procedure (The sixth order conformally invariant operator in $d=6$ was obtained in \cite{MMK} from cohomological consideration).
We performed the full calculation  inserting (\ref{31}) with   $k=3$ in (\ref{25}) and have found full agreement with the formula (56) of \cite{MT}. In this article, to avoid cumbersome formulas, we will trace only the derivative independent term linear in $\Delta_{(3)}$.
First of all we see from (\ref{25}) and (\ref{30}) the relation
\begin{eqnarray}\label{41}
    \Box_{A}\ell^{4}t^{\Delta_{(k)}-4}f_{(k)}(\rho,x)&=&\ell^{6}t^{\Delta_{(k)}-6}
    \left[\Box+(4-\Delta_{(k)})h^{(1)}\right.\nonumber\\&+&\left.4(5-k)\partial_{\rho} +O(\rho)\right]f_{(k)}(\rho,x)\,.
\end{eqnarray}
Then it is easy to see that the relevant terms in (\ref{31}) are only two derivative free expressions with the $\ell^{-4}$ in front. Now because both derivative free terms in (\ref{31}) are already with a $\Delta_{(k)}$ factor, the operator (\ref{41}) contributes only as $4h^{(1)}+8\partial_{\rho}$ if $k=3$ and we have to just multiply the derivative free part of the second line in (\ref{31}) (it is just $-\frac{2\Delta_{(3)}}{\ell^{4}}h^{(1)2}$ for k=3) by $4h^{(1)}$ and add it to the third line of (\ref{31}) with factor $8$ instead of the $\rho$. So finally we have
\begin{eqnarray}
  && \Box^{3}_{A}\left[t^{\Delta_{(3)}}\varphi_{(k)}(x)\right]|_{\rho=0}
  = \ell^{6}t^{\Delta_{(3)}-6}\hat{\mathcal{L}}_{(3)}\varphi_{(3)}(x)=\ell^{6}
  t^{\Delta_{(3)}-6}\Big\{\Box^{3}+ \dots\dots \nonumber\\ && +\frac{8\Delta_{(3)}}{\ell^{6}}\left[8h^{(1)ij}h^{(2)}_{ij}
  -4h^{(1)ij}h^{(1)}_{jk}h^{(1)k}_{i}+3 h^{(1)}
  h^{(1)ij}h^{(1)}_{ij}-h^{(1)3}\right]\Big\}\varphi_{(3)}(x) .\quad\label{42}
\end{eqnarray}
Now using again (\ref{12}) and (\ref{13}) we see that
\begin{eqnarray}
  && \mathbf{a}_{(3)}=-8\left[J^{3}-3K^{ij}K_{ij}J+
  2K^{ij}K_{jn}K^{n}_{j}-\frac{2}{d-4}K^{ij}B_{ij}\right].\label{43}
\end{eqnarray}
We see again that this part coincides with the socalled
"holographic" anomaly \cite{HS} in 6 dimensions written in
general spacetime dimension $d$ ( see  also \cite{OA}). The important property of the holographic anomaly is that it is a special combination of the Euler density
with three other  Weyl invariants \cite{Deser:1996na},\cite{Bastianelli:2000hi}  which
reduce the topological part of the anomaly to the expression (\ref{43})
, which is zero for the Ricci flat metric (see  \cite{Boul2} for recent results on purely algebraic considerations of the general structure of the Weyl anomaly in arbitrary $d$ ).

\section{The ambient space, PBH diffeomorphisms  and Ricci gauging}
In this section we consider an ambient space origin of another method of construction of $d$ dimensional local conformal invariants. This is the so-alled Ricci gauging proposed by A.~Iorio, L.~O'Raifeartaigh, I.~Sachs and C.~Wiesendanger in \cite{oraf}. Ricci gauging is very effective when we start from a scale invariant matter field Lagrangian and want to generalize it to a local Weyl or conformal invariant Lagrangian.
The prescription developed in \cite{oraf} consists of two steps
\begin{enumerate}
\item First of all we have to perform Weyl gauging by introduction of the corresponding Weyl gauge field $A_{i}(x)$. For the scalar field it looks like
   \begin{eqnarray}
     &&  \partial_{i}\varphi_{(k)}(x)\rightarrow D_{i}\varphi_{(k)}(x)= (\partial_{i}-\Delta_{(k)}A_{i}(x))\varphi_{(k)}(x) ,\label{44}\\
     && \delta A_{i}(x)=\partial_{i}\sigma(x) ,\quad \delta D_{i}\varphi_{(k)}(x)=\Delta_{(k)}D_{i}\varphi_{(k)}(x),\,\label{45}
   \end{eqnarray}
   with the additional "pure gauge" conditions $\nabla_{i}A_{j}=\nabla_{j}A_{i}$ for elimination of the  self invariant combinations of $A_{i}$ constructed from the field strength $F_{ij}=\partial_{[i}A_{j]}$ .
   \item After Weyl gauging the actions with a conformally invariant flat space limit (scale invariant) contain  the field $A_{i}$ only in the combinations
       \begin{eqnarray}
         && \Omega_{ij}[A]=\nabla_{i}A_{j}(x)+A_{i}A_{j}-\frac{g_{ij}}{2}g^{kl}A_{k}A_{l} ,\quad \delta \Omega_{ij}[A]=\nabla_{i}\partial_{j}\sigma(x) \,,\label{46}\\
         && \Omega[A]=g^{ik}\nabla_{i}A_{k}(x)+\frac{d-2}{2}g^{kl}A_{k}A_{l} ,\quad \delta \Omega[A]=\Box \sigma(x)  \,,\label{47}
       \end{eqnarray}
   and therefore can be replaced by
   \begin{equation}\label{48}
    K_{ij}=\Omega_{ij}[A]\quad \textnormal{and}\quad  J=\Omega[A]\,.
   \end{equation}
    The authors of \cite{oraf} called  this procedure \emph{Ricci gauging}.
   \end{enumerate}

To understand this Ricci gauging on the level of $d+2$ dimensional \emph{gauged} ambient space of Fefferman and Graham we turn first to the idea of PBH diffeomorphisms \cite{PBH} of the higher dimensional spaces, which reduce to conformal transformations on the lower dimensional boundary or embedded subspace. Actually the PBH transformations can be defined as higher dimensional diffeomorphisms which leave the form of the higher dimensional metric invariant.
The PBH transformations for the bulk metric (\ref{9}) are constructed and analyzed in \cite{Schwim} and \cite{S}. For the $d+2$ dimensional ambient metric (\ref{1}) PBH diffeomorphisms are considered in \cite{OA}. The existence of such a transformations is another reason why the reparametrization invariant powers of the Laplacian in ambient space reduce to the Weyl invariant
operators in $d$ dimensional space as considered in the previous section.
Following \cite{OA} we define PBH transformations of (\ref{1}) as diffeomorphisms (Lie derivative along the vector $\zeta^{\mu}(t,\rho,x)$)
\begin{eqnarray}
    \delta g^{A}_{\mu\nu}(x^{\mu})&=&\mathcal{L}_{\zeta(t,\rho,x)}g^{A}_{\mu\nu}(t,\rho,x)
    =\zeta^{\lambda}(t,\rho,x)\partial_{\lambda}g^{A}_{\mu\nu}(t,\rho,x)\nonumber\\
    &+&g^{A}_{\mu\lambda}(t,\rho,x)\partial_{\nu}\zeta^{\lambda}(t,\rho,x)
    +g^{A}_{\nu\lambda}(t,\rho,x)\partial_{\mu}\zeta^{\lambda}(t,\rho,x) ,\label{49}
\end{eqnarray}
satisfying the conditions
\begin{equation}\label{50}
    \delta g^{A}_{tt}(t,\rho,x)=\delta g^{A}_{t\rho}(t,\rho,x)
    =\delta g^{A}_{\rho\rho}(t,\rho,x)=\delta g^{A}_{ti}(t,\rho,x)=g^{A}_{\rho i}(t,\rho,x)=0\,.
\end{equation}
The corresponding infinitesimal PBH transformations are \cite{Schwim},\cite{OA}
\begin{eqnarray}
  && \zeta^{t}(t,\rho,x)=t\sigma(x) \,,\label{51}\\
  && \zeta^{\rho}(t,\rho,x)=-2\rho\sigma(x)\,,\label{52}\\
  && \zeta^{i}(t,\rho,x)=\zeta^{i}(x,\rho) \,,\quad h_{ij}(\rho ,x)\partial_{\rho}\zeta^{i}(\rho,x)=\frac{\ell^{2}}{2} \partial_{i}\sigma(x)\,,\label{53}\\
  && \delta h_{ij}(\rho,x)=2\sigma(x)(1-\rho\partial_{\rho})h_{ij}(\rho,x)
  +\mathcal{L}_{\zeta(\rho,x)}h_{ij}(\rho,x) \,.\label{54}
\end{eqnarray}
We see that PBH transformations depend on two free parameters $\sigma(x)$ and $\zeta^{i}(x)=\zeta^{i}(0,x)$ corresponding to the local Weyl and local diffeomorphisms of the boundary metric $g_{ij}(x)=h_{ij}(0,x)$. All other terms $n!\zeta^{(n)i}(x)=\frac{\partial^{n}}{\partial_{\rho}^{n}}\zeta(\rho,x)|_{\rho=0}$ of the $\rho$ expansion of the $\zeta^{i}(\rho,x)$ are expressed through $\sigma(x)$ according to the relation (\ref{53}). This dependence fixes the special unhomogeneous  forms of the Weyl transformations of the FG coefficients,  which is in full agreement with the  direct solution (\ref{11})-(\ref{13}) of the corresponding equations (\ref{3}) or (\ref{11}) (see \cite{Schwim} for details).

To include the Weyl gauge field $A_{i}(x)$ in this game and find an ambient space description of the Ricci gauging we introduce a generalized $d+2$ dimensional gauged ambient space with the following metric
\begin{equation}\label{55}
    ds^{2}_{GA}=\frac{t^{2}}{\ell^{2}}\left[h_{ij}(\rho,x)+\rho \ell^{2}A_{i}(x)A_{j}(x)\right]dx^{i}dx^{j}-\rho dt^{2} -t\left[dt+tA_{i}(x)dx^{i}\right]d\rho\,.
\end{equation}
Then we consider corresponding $d+2$ dimensional diffeomorphisms conserving the form of (\ref{55})
\begin{equation}\label{56}
  \delta g^{GA}_{tt}(t,\rho,x)=\delta g^{GA}_{t\rho}(t,\rho,x)
    =\delta g^{GA}_{\rho\rho}(t,\rho,x)=\delta g^{GA}_{ti}(t,\rho,x)=0\,,
\end{equation}
and giving for $A_{i}(x)$  a gauge transformation with the Weyl parameter $\sigma(x)$ (\ref{45}). The corresponding solution gives for new PBH transformations
\begin{eqnarray}
  && \zeta^{t}(t,\rho,x)=t\sigma(x) \,,\label{57}\\
  && \zeta^{\rho}(t,\rho,x)=-2\rho\sigma(x)\,,\label{58}\\
  && \zeta^{i}(t,\rho,x)=\zeta^{i}(x) \,,\label{59}\\
  && \delta h_{ij}(\rho,x)=2\sigma(x)(1-\rho\partial_{\rho})h_{ij}(\rho,x)
  +\mathcal{L}_{\zeta(x)}h_{ij}(\rho,x) \,,\label{60}\\
  &&\delta A_{i}(x)=\partial_{i}\sigma(x) +\mathcal{L}_{\zeta(x)}A_{i}(x)\,.\label{61}
\end{eqnarray}
Comparing with (\ref{51})-(\ref{54}) we see that we were lucky with the ansatz (\ref{55}) to restore the Weyl part of the PBH transformation with the proper gauge transformation for $A_{i}(x)$.
The only difference that we have here is the $\rho$-independence of the bulk diffeomorphisms $\zeta^{i}(x)$ and correspondingly the absence of the  condition (\ref{53}).
It is a price for the additional gauge field transformation (\ref{61}). However, this difference is very essential for the FG expansion. Putting $\zeta^{i}(x)=0$ we get from (\ref{60}) for pure Weyl transformations of the FG coefficients $n!h^{(n)}_{ij}(x)$ only the homogeneous parts
\begin{eqnarray}
  && \delta g_{ij}(x)=2\sigma(x)g_{ij}(x) \,,\label{62}\\
  && \delta h_{ij}^{(1)}(x)=0 \,,\label{63}\\
  && \delta h_{ij}^{(2)}(x)=-2\sigma(x)h^{(2)}_{ij}(x)\,.\label{64}
\end{eqnarray}
 So it seems really as a Weyl gauged version of the FG expansion.
For making the final check of this assertion we turn now to the Ricci flatness condition for the gauged ambient metric (\ref{55}). Inverting the metric (\ref{55}) we obtain
\begin{eqnarray}
&&\left(
  \begin{array}{ccc}
   \ell^{2}A^{2} & -\frac{2\gamma}{t} & -\frac{\ell^{2}}{t}A^{j}\\
    -\frac{2\gamma}{t} & \frac{4\rho\gamma}{t^{2}} & \frac{2\rho\ell^{2}}{t^{2}} A^{j}\\
     -\frac{\ell^{2}}{t}A^{i} & \frac{2\rho\ell^{2}}{t^{2}} A^{i} & \frac{\ell^{2}}{t^{2}}h^{ij} \\
  \end{array}
\right)\,,\label{65}\end{eqnarray}
where
 \begin{eqnarray}
   && \gamma=1+\rho\ell^{2}A^{2} , \quad A^{2}(\rho,x)=h^{nm}(\rho,x)A_{n}(x)A_{m}(x)\,,\label{66}\\
   && A^{i}(\rho,x)=h^{ik}(\rho,x)A_{k}(x)\,.
 \label{67}
 \end{eqnarray}
 Then  the calculation of the Christoffel symbols and Ricci tensor became straightforward if we admit the condition $F_{ij}=0$ .
After a long calculation we see that the first four equations
\begin{eqnarray}
  && R^{GA}_{it} = R^{GA}_{\rho t}= R^{GA}_{tt}\equiv 0 \,,\label{68}\\
  &&R^{GA}_{\rho\rho} =\frac{1}{2}\left[h^{kl}h''_{kl}
  -\frac{1}{2}h^{ij}h'_{jk}h^{kl}h'_{li}\right]=0 \,,\label{69}
\end{eqnarray}
are the same as in the usual ambient space. But the last two undergo a change
\begin{eqnarray}
   R^{GA}_{i\rho}&=&\frac{1}{2}h^{kl}\left[\nabla^{(h)}_{i}
  h'_{kl}-\nabla^{(h)}_{k}h'_{il}\right] + \frac{1}{2}h^{kl}h'_{kl}A_{i}+\frac{d-2}{2}h'_{ik}h^{kl}A_{l}-\rho h''_{ik}h^{kl}A_{l}=0\nonumber\\&&\hspace{8cm}\,,\label{70}\\
  \ell^{2}R^{GA}_{ij}&=&\ell^{2}R_{ij}[h]-(d-2)h'_{ij} - \gamma h^{kl}h'_{kl}h_{ij}+
  \rho\gamma\left[2h''_{ij}-2h'_{il}h^{lm}h'_{mj}
  +h^{kl}h'_{kl}h'_{ij}\right]\nonumber\\
  &-&(d-2)(\nabla^{(h)}_{i}A_{j}+A_{i}A_{j}-A^{2}h_{ij})-h_{ij}\nabla^{(h)}_{k}A^{k}\nonumber\\
  &+&\rho[h^{kl}h'_{kl}\nabla^{(h)}_{i}A_{j}-(d-4)A^{2}h'_{ij}
  -2A^{k}(h'_{ik}A_{j}+h'_{jk}A_{i})\nonumber\\
  &-& h^{kl}(h'_{ki}\nabla^{(h)}_{l}A_{j}+h'_{kj}\nabla^{(h)}_{l}A_{i})+\nabla^{(h)}_{k}(h'_{ij}A^{k})+2\rho h'_{ik}A^{k}h'_{jl}A^{l}]=0\,.\label{71}
\end{eqnarray}
Then inserting in (\ref{71}) $\rho=0$ we obtain instead of (\ref{12}) the following solution for the first coefficient of the FG expansion
\begin{eqnarray}
  \frac{1}{\ell^{2}}h^{(1)}_{ij}(x)&=& K_{ij}-\nabla_{i}A_{j}-A_{i}A_{j}
  +\frac{1}{2}g_{ij}A_{k}A_{l}g^{kl}\nonumber\\
  &=& K_{ij}-\Omega_{ij}[A] \,,\label{72}\\
 \frac{1}{\ell^{2}}h^{(1)}(x)&=& J-\Omega[A] \,.\label{73}
\end{eqnarray}
So we see that (\ref{72}) is Weyl invariant which is in agreement with the PBH transformation (\ref{63}). On the other hand we see that Ricci gauging  leads to a \emph{trivialization of the Fefferman-Gracham expansion}. Indeed the Ricci gauging condition (\ref{48}) means
\begin{equation}\label{74}
    h^{(1)}_{ij}\equiv 0 .
\end{equation}
Moreover because equations (\ref{69})-(\ref{71}) express recursively each next $h^{(n)}_{ij}$  through the nonzero powers of previous ones  we can conclude that all higher $h^{(n)}_{ij}$ coefficients of the FG expansion are trivialized after imposing the Ricci gauging condition.
The final conclusion which we can make now is the following:
\emph{The FG expansion for a gauged ambient metric (\ref{55}) can be obtained from the usual expansion for (\ref{1}) by  the Weyl gauging}.
For example we can easily guess the next coefficient
\begin{equation}\label{75}
    h^{(2)}_{ij}(x)=\frac{\ell^{4}}{4}\left\{\frac{\tilde{B}_{ij}}{d-4}+(K^{m}_{i}
    -\Omega^{m}_{i}[A])(K_{mj}-\Omega_{mj}[A])\right\} ,
\end{equation}
where
\begin{equation}\label{76}
    \tilde{B}_{ij}=B_{ij}-(d-4)A^{k}(C_{kij}+C_{kji})-(d-4)A^{k}A_{l}W^{\quad l}_{kij}
\end{equation}
is the Weyl gauged Bach tensor.
\section*{Conclusion}
We have proved our assertion concerning the connection between the
hierarchy of conformally coupled scalars with the dimensions
$\Delta_{(k)}$  and the FG ambient space method of construction of the conformal invariants \cite{FG}. This consideration explains in a natural way the origin of the holographic structure emerging in the study of the higher derivative invariant operators in \cite{MT} obtained using the direct Noether procedure.
In the last part of these notes we evaluated the $d+2$ dimensional ambient space origin of the Ricci gauging \cite{oraf} which is another powerful method for building local conformal invariant Lagrangians.

\subsection*{Acknowledgements}
\quad R. Manvelyan is very grateful to D.H. Tchrakian and W. R\"uhl for numerous thoughtful
discussions on this subject. This work is partially supported by INTAS grant
\#03-51-6346.

\end{document}